\documentclass[12pt]{article} \textheight 23cm
\usepackage{graphicx}
\usepackage{amsmath}
\usepackage{color}
\begin{document} \begin{center}
{\large Identifying `Island-Mainland' phase transition using the Euler number    }\\ \vskip 0.5cm
Tajkera Khatun$^1$, Tapati Dutta$^2$ and Sujata Tarafdar$^{3*}$\\
\vskip 0.5cm

$^1$ Physics Department, Charuchandra College, Kolkata 700029, India\\
$^2$ Physics Department, St. Xavier's College, Kolkata 700016, India\\
$^3$ Condensed Matter Physics Research Centre, Physics Department, Jadavpur University, Kolkata 700032, India\\

$^*$ Corresponding author, email: sujata$\_$tarafdar@hotmail.com, Phone: +913324146666(Extn. 2760), Fax: +913324138917
\end{center}
\vskip .5cm
\noindent {\bf Abstract}\\ 
In the present communication we describe the \textit{Island-Mainland} transition, occurring in a square lattice, when black squares are randomly dropped on a white background, with increasing concentration $p$. Initially clusters of black squares (islands) are observed on the connected white background forming a `sea'. But as concentration of black sites increases, at some point the background break up. As deposition of black squares continues the system passes through a `mixed phase' (MP) where neither the black nor white regions are fully connected. With further increase in concentration of black squares, the black squares form a fully connected background, which now becomes the `mainland' with isolated pockets of white (lakes). So, as `p' goes from 0 to 1,  the system undergoes the following sequence of transitions \textit{Islands in sea}(IS)$\rightarrow$\textit{Mixed Phase}(MP)$\rightarrow$\textit{Lakes in mainland}(LM). We show that the Euler number $\chi$, defined as the difference between number of white clusters and number of black clusters, is maximum at the IS$\rightarrow$MP transition and minimum at the MP$\rightarrow$LM crossover. 
 We show that the phenomenon can be related to experimental observations in several physical systems. 
\vskip .3cm

We describe here, a transition sequence often observed in many physical systems which we designate as the \textit{Island-Mainland transition}. The subject of phase transitions is one of the most widely studied in condensed matter physics \cite{stanley}, but there still remain observations waiting to be properly characterized and classified. The case we present here is a transition related to, but distinct from the well-studied percolation phase transition \cite{stauffer}. Supposing one starts with a square lattice which has all square cells white and randomly colors a fraction $p$ of the cells black, increasing $p$ gradually from 0 to 1. Initially there would be scattered black cells, or clusters of black cells, which we term  \textit{islands} in a connected \textit{sea} of white cells. Here we define connected clusters as a group of cells of same color which share edges or corners. 

Let us denote the number of connected black \textit{clusters} by $N_B$ and the number of connected white \textit{clusters} by $N_W$. When $p = 0$, $N_B = 0$ and $N_W = 1$.
As $p$ increases from 0, $N_B$ increases upto the stage where black clusters are small and isolated, but later as black islands start to get connected with each other $N_B$ starts to fall. Meanwhile the ocean of white clusters starts to break up into smaller seas or lakes, so $N_W$ increases too. As $N_W$ deviates from a value very close to 1 to non-zero values, the system enters a \textit{mixed phase} (MP) with neither black sites nor white sites forming a fully connected background.

However as $p$ becomes larger and larger approaching 1, $N_W$ must again decrease and ultimately reach 0 when $p = 1$. So after a certain $p$,  the MP phase crosses over to a phase where isolated white \textit{lakes} are scattered randomly on a  black \textit{land mass} or \textit{mainland}. We call this the MP$\rightarrow$LM transition.

We show that this transition sequence represents a concept different from the \textit{\textit{percolation}} transition, the percolation threshold is the point $p_c$ where the black squares first form a `connected cluster' spanning the system. If two conducting bars were placed along the upper and lower sides of the square and connected to a voltage source, $p_c$ is the point where a non-zero current passes through the sample. To analyze the evolution of the complex morphology of the system and identify the IS$\rightarrow$MP$\rightarrow$LM transition sequence, we use the topological concept of the Euler number $\chi$ \cite{euler}, which is simply 
\begin{equation}
\chi(p) = N_B(p) - N_W(p)
\end{equation}
Obviously
\begin{equation}
\chi(0) = -1
\end{equation}
and 
\begin{equation}
\chi(1) = 1
\end{equation}

\begin{figure}[h]
\begin{center}
\includegraphics[width=12.0cm,angle=0]{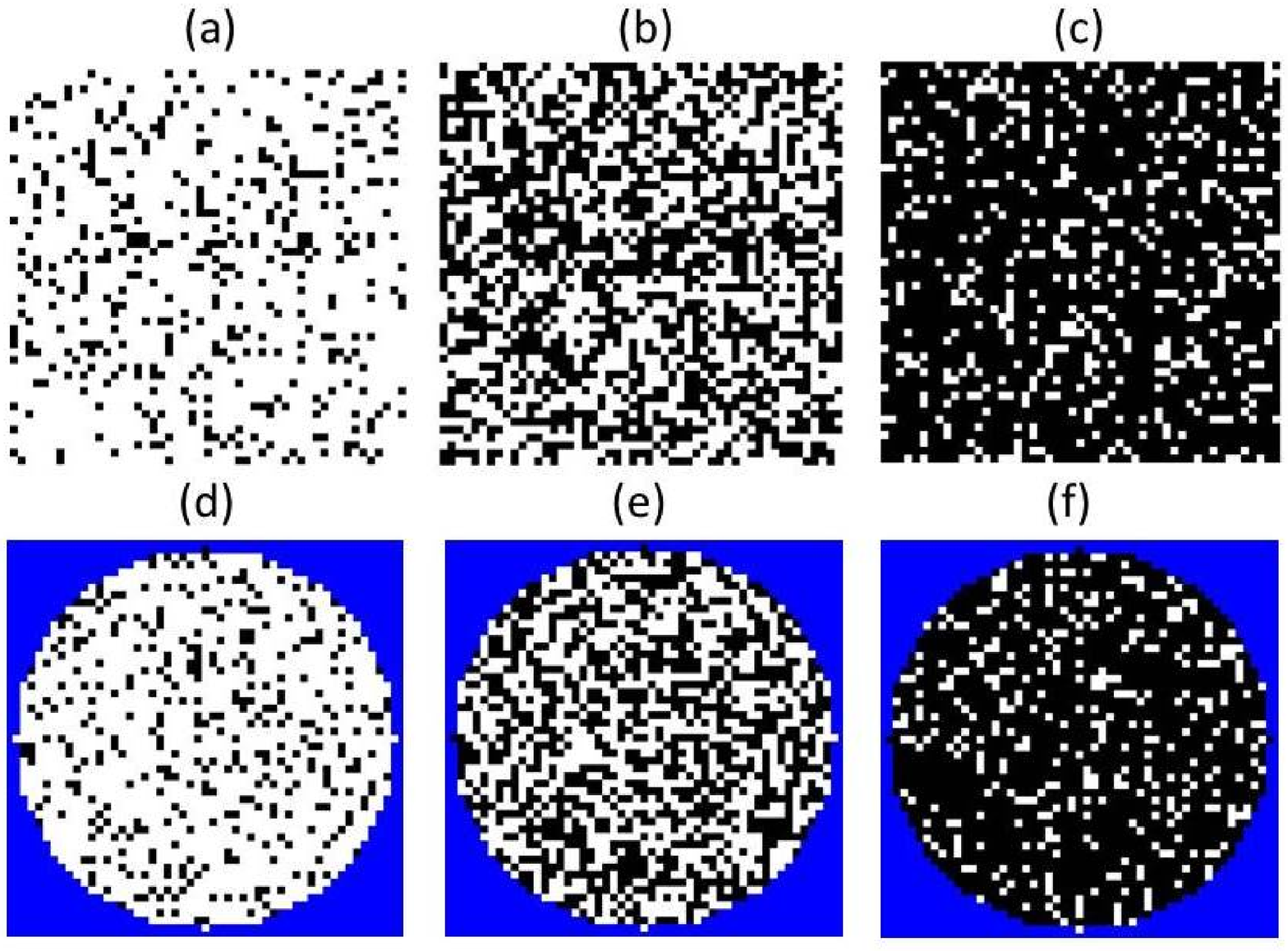}
\end{center}
\caption{Random deposition patterns of a concentration $p$ of black squares on an initially white square lattice are shown for $p$ = 0.17, 050 and 0.83. Results for a system with a square boundary of size $51\times 51$ averaged over 100 configurations are shown in (a), (b) and (c). Results for a system with a circular boundary of radius 25 are shown in (d), (e) and (f).}
\label{sqr-circ}
\end{figure}

\begin{figure}[h]
\begin{center}
\includegraphics[width=12.0cm,angle=0]{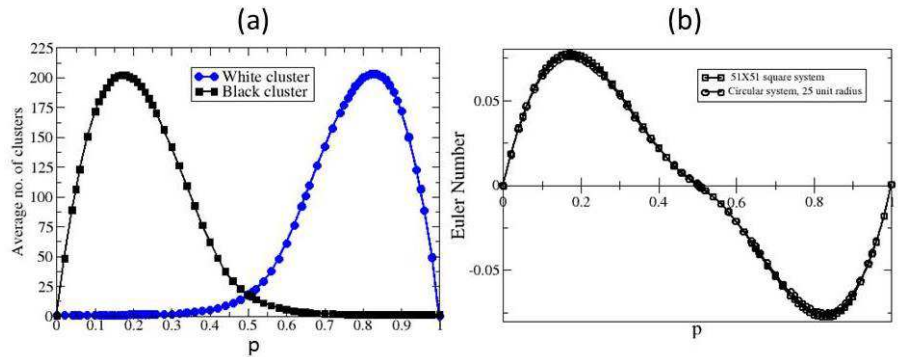}
\end{center}
\caption{(a) Variation of the number of black clusters, $N_B$ with $p$ for the square system are shown as black squares. Number of white clusters, $N_W$ with $p$ for the square system are shown as blue solid circles. (b) variation of the Euler number $\chi$ with $p$ is shown. Results for the square and circular systems overlap when normalized by the area of the system.}
\label{chi}
\end{figure}

We simulate the situation described above for a system of size $L\times L$ on a square lattice using a Fortran program. We increase $p$ in steps and calculate the evolution of $\chi(p)$. Since the system is stochastic, we average the result over $N_{conf}$ configurations. The appearance of the system for several $p$ values for a square lattice with both square and circular boundaries is shown in fig(\ref{sqr-circ}). The graph of $\chi$ plotted against $p$  averaged over $N_{conf}$=100 configurations, on a 51$\times$51 system is shown in fig(\ref{chi}a). We see that $\chi$ first reaches a maximum at $p=0.17$, then falls, has a slight inflection when crossing the X-axis,  and decreases further until it reaches a minimum at $p=0.83$. Following this it climbs up, finally reaching the value $\chi(p)=1$ at $p=1$. 
The simulation was repeated for several system sizes with similar results.

Fig(\ref{chi}a) shows that $N_W$ is very near 1 from $p=0$ up to $p=0.17$ where $N_B$ reaches a maximum. Fig(\ref{chi}b) shows that $\chi$ reaches a maximum here, we identify this point as $p_{c1}$ the IS$\rightarrow$MP transition. The graphs for $N_B$ and $N_W$ are mirror images of each other, but the slopes to the right and left of each peak are not identical. The minimum of $\chi$ can be identified with the MP$\rightarrow$LM transition after which the black clusters form a fully connected system.

The percolation threshold ($p_c$) has been calculated by implementing the Hoshen-Kopelman algorithm  \cite{hoshen} considering the cluster to include second nearest neighbors. The threshold in this case is at $p=0.39$, when the first system spanning black cluster appears with $p$ increasing from 0. The corresponding threshold for the white spanning cluster when $p$ decreases from 1 is $p=0.61$.

  We illustrate the greater generality of the IM transition, by working out the whole process on a square lattice with a circular boundary (fig(\ref{sqr-circ})d-f). If the system is confined by a circular boundary, the concept of the percolation transition is no longer meaningful. Here we cannot define a system spanning `infinite'  cluster, stretching from one side to another. However, the IM transition remains equally significant as for the square system with straight boundaries. $\chi(p)$ calculated for the circular system fig(\ref{chi}b) shows that $\chi_{max}$ and $\chi_{min}$ occur at the same value of $p_{c1}$ and $p_{c2}$ as for the square system. Further, the curve for $\chi$, when normalized by the total area of the system i.e. $side^2$ for the square and $\pi \times radius^2$ for the circular, match each other exactly as shown in fig(\ref{chi}b). The inflection point represents $\chi$=0, which occurs from symmetry at $p$=0.5, here $N_B=N_W$.
  
\begin{figure}[h]
\begin{center}
\includegraphics[width=12.0cm,angle=0]{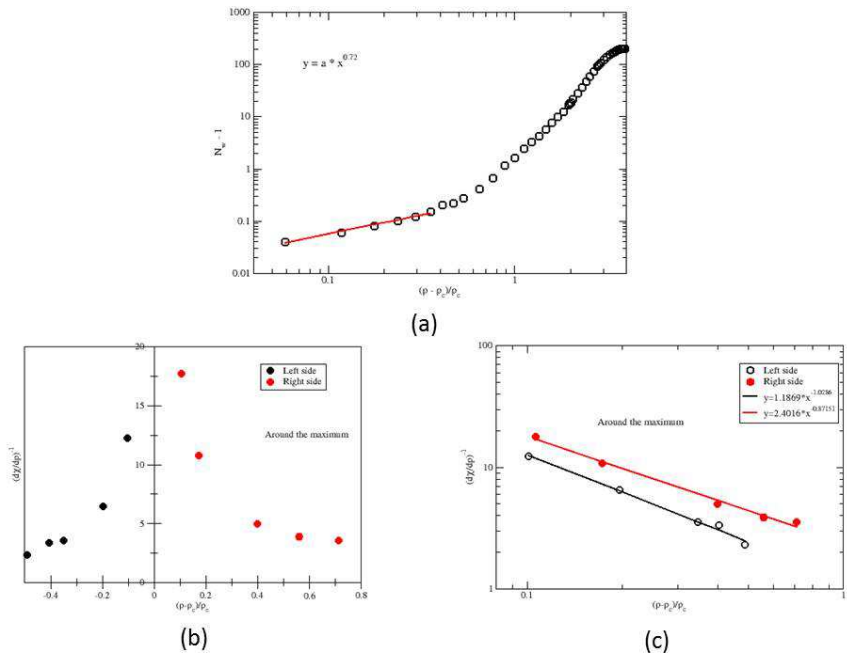}
\end{center}
\caption{(a) the order parameter $Q(p)$ against $p_{red}$, as $p$ approaches the transition $p_{c1}$ from above following a power law with exponent 0.72. (b) $|\partial\chi/\partial p|$ against $p_{red}$ showing divergence near $p_{c1}$. (c) A log-log plot of the data in (b) shows power law variation with exponents -1.02 and -0.87 for $p<p_{c1}$ and $p>p_{c1}$ respectively. }
\label{pwr-law}
\end{figure}

 The well-known results for critical exponents in phase transitions \cite{stanley} prompts us to look for similar characteristics in the IS$\rightarrow$MP and MP$\rightarrow$LM transitions observed here. Since $N_W$=1 in the IS phase and non-zero in the other phases, we define the order parameter for the IS$\rightarrow$MP transition as 
\begin{equation}
Q(p)= N_W - 1
\end{equation} 
and plot this as a function reduced $p$ defined as
\begin{equation}
p_{red} = |(p-p_{c1})|/p_{c1}
\end{equation}
A similar behavior is  found from symmetry at $p_{c2}$, the other transition point. We show in fig\ref{pwr-law} that near $p_{c1}$, $Q(p)$ approaches 0 for decreasing $p$ with a power-law having exponent 0.72 as evident from the log-log plot in figure(\ref{pwr-law}a).

Another parameter we find of interest is the curve for the inverse of the derivative of the Euler number as function of $p$, i.e. $(\partial\chi/\partial p)^{-1}$, which approaches $\pm \infty$ at $p_{c1}$ and $p_{c2}$. The magnitude of these derivatives are shown in figure(\ref{pwr-law}b) near $p_{c1}$ as a function of $p_{red}$. Similar symmetric curves can be drawn near $p_{c2}$. When plotted on a logarithmic scale, these results show  power-law behavior with exponents -1.03 for $p<p_{c1}$ and -0.87 for $p>p_{c1}$. These results indicate scale invariance and critical behavior near the transitions studied here. Work on the variation of the correlation length is in progress.
We now discuss the relevance of these transitions observed in two different experimental situations observed in our lab.

Two recent experiments have demonstrated that the MP$\rightarrow$LM transition identified by $\chi_{min}$ plays a significant role in evolving complex systems. The experiments and relevant results are as follows.
\begin{enumerate}
\item{
Dutta Choudhury et al. \cite{mdc} studied crystal growth during evaporation of a sessile droplet of complex fluid containing gelatin, water and sodium chloride. Videos of the drying process \cite{video} showed a transition between formation of faceted large crystals in early stages of drying, to intricate dendritic patterns in later stages. From the video it appears that as evaporation proceeds, voids form in the continuous film of fluid and at some point, the film breaks up into small isolated fluid blobs. This crossover is similar to the MP$\rightarrow$LM transition proposed here. A simulation of the evaporation and crystal growth was carried out by Dutta Chowdhury et al. \cite{mdc}. The Euler number for the fluid and void clusters (which are analogous to the white and black clusters respectively, in the present paper ) was calculated and the transition from faceted to dendritic growth was switched on  when $\chi$ reached a minimum. The simulation agreed very closely with the experimentally observed time development of the crystal growth pattern.}
\item{
The other experiment studies desiccation crack formation in layers of clay \cite{khatun}. The crack pattern depends on the thickness of the clay layer. There is a critical cracking thickness, below which cracks do not form at all. For a  thin layer, just thicker than the critical, isolated cracks form randomly over the whole layer. As the thickness is increased further, cracks start to form a connected network. Regions of the clay layer now  appear which are bounded by cracks on all sides, these are termed \textit{peds}. With increase of the clay layer thickness, the average distance between cracks increases and the number of peds decreases. We may thus identify the number of peds with $N_W$ and number of cracks as $N_B$. The Euler number, as function of layer thickness was calculated in this way and showed a minimum at the thickness where the crack network becomes fully connected.}
 \end{enumerate}
To conclude we present a preliminary description of a transition we name as the `Island-Mainland' transition. This is seen to be a physically significant phenomenon, observed in simulated as well as real systems. It needs to be studied in further detail and properly characterized to reveal its full richness.

 \newpage

\end{document}